\documentclass[10pt,conference]{IEEEtran}

\usepackage{flushend}

\usepackage{amsmath,amssymb,amsthm,amsfonts,latexsym,bbm,xspace,graphicx,float,mathtools}
\usepackage[backref,colorlinks,citecolor=blue,bookmarks=true]{hyperref}
\usepackage[letterpaper,margin=1in]{geometry}
\usepackage{color}
\usepackage{algorithm}
\usepackage{algorithmic}

\newtheorem{theorem}{Theorem}

\newtheorem{lemma}{Lemma}
\newtheorem{conjecture}{Conjecture}
\newtheorem{claim}{Claim}

\newtheorem{definition}{Definition}

\newtheorem{remark}{Remark}

\newenvironment{prevproof}[2]{\noindent {\em {Proof of {#1}~\ref{#2}:}}}{$\Box$\vskip \belowdisplayskip}

\IEEEoverridecommandlockouts

\begin{document}

\title{A Counter-Example to Karlin's Strong Conjecture for Fictitious Play}   
\author{Constantinos Daskalakis$\text{}^{\dag}$\thanks{$\text{}^{\dag}$Supported by a Sloan Foundation fellowship, a Microsoft Research faculty fellowship and NSF Award CCF-0953960 (CAREER) and CCF-1101491.}\\EECS, MIT\\costis@mit.edu \and Qinxuan Pan$\text{}^{\ddag}$\thanks{$\text{}^{\ddag}$Supported by ONR grant N00014-12-1-0999.}\\EECS, MIT\\qinxuan@mit.edu}
\date{\today}   
\addtocounter{page}{-1}
     
\maketitle

\begin{abstract}
Fictitious play is a natural dynamic for equilibrium play in zero-sum games, proposed by Brown~\cite{Brown49}, and shown to converge by Robinson~\cite{robinson1951iterative}. Samuel Karlin conjectured in 1959 that fictitious play converges at rate $O(t^{-\frac{1}{2}})$ with respect to the number of steps $t$. We disprove this conjecture by showing that, when the payoff matrix of the row player is the $n \times n$ identity matrix, fictitious play may converge (for some tie-breaking) at rate as slow as $\Omega(t^{-\frac{1}{n}})$.

\end{abstract}
\thispagestyle{empty}

\section{Introduction}\label{sec-introduction}

Von Neumann's MinMax theorem for two-person zero-sum games marked the birth of Game Theory~\cite{vN}, and is intimately related to the development of linear programming. Given a payoff matrix $A$, whose $ij$-th entry specifies how much the column player playing $j$ pays the row player playing $i$, the theorem states that
\begin{align*}
\max_{x} \min_{y} x^{\rm T} A y=\min_{y} \max_{x} x^{\rm T} A y, 
\end{align*}
where $x$, $y$ range over randomized/mixed strategies for the row and column player respectively. In other words, there exists a unique value $z \in \mathbb{R}$ and a pair of mixed strategies $\hat{x}$ and $\hat{y}$ such that:
\begin{align}
\min_{y} \hat{x}^{\rm T} A y = z = \max_{x} x^{\rm T} A \hat{y}.\label{eq: minmax equilibrium}
\end{align}

Dantzig and von Neumann observed that the MinMax theorem is implied by strong linear programming duality~\cite{Dantzig63,Interview_with_Dantzig}. Dantzig also provided a candidate construction for the opposite implication~\cite{Dantzig63}, and this was also established some decades later~\cite{Adler13}.

Ultimately, the MinMax theorem provides a very sharp prediction in two-player zero-sum games. It shows that there is a unique value $z$ and a pair of strategies $\hat{x}$ and $\hat{y}$ such that, by playing $\hat{x}$ the row player can guarantee himself expected payoff of $z$ regardless of what strategy the column player adopts, and such that, by playing $\hat{y}$, the column player can guarantee herself expected payoff of $-z$ regardless of what strategy the row player adopts. In particular, $(\hat{x},\hat{y})$ comprise a Nash equilibrium of the game, with expected payoff $z$ for the row player and $-z$ for the column. Moreover, $\hat{x}$, $\hat{y}$ and $z$ can be computed in polynomial time with linear programming. This type of crisp prediction is rather rare in Game Theory. According to Aumann, zero-sum games are ``one of the few areas in game theory, and indeed in the social sciences, where a fairly sharp, unique prediction is made''~\cite{aumann1987game}.

Shortly after the proof of the MinMax theorem and the development of linear programming, G.~W.~Brown proposed {\em fictitious play} as an iterative procedure for solving a zero-sum game, or equivalently a linear program~\cite{Brown49,brown1951iterative}. The procedure proceeds in steps in which players choose a pure strategy best response to their opponent's empirical mixed strategy up until that step. Let us describe it a bit more formally (we focus on the {\em simultaneous} version, but our results also hold for the {\em asynchronous} version, where the players' moves alternate): At every step $t$, the row player chooses some row $i_t$ and the column player chooses some column $j_t$. At $t=1$, the choices are arbitrary. At $t+1>1$, the players calculate the \textit{empirical mixed strategies} of their opponents in previous steps, namely\footnote{We use $e_i$ to denote the column vector with $i$-th component $1$ and all other components $0$. The dimension of $e_i$ is always implied by the context; it is $m$ when describing row player strategies and $n$ when describing column player strategies.}
\begin{equation*}
\begin{aligned}
x(t) & = {1 \over t}\sum_{\tau \leq t} e_{i_{\tau}}, \\
y(t) & = {1 \over t}\sum_{\tau \leq t} e_{j_{\tau}}.
\end{aligned}
\end{equation*}
Then, the row player chooses an arbitrary best response $i_{t+1}$ to $y(t)$ and the column player chooses an arbitrary best response $j_{t+1}$ to $x(t)$, namely
\begin{equation}\label{eq:ties}
\begin{aligned}
i_{t+1} & \in \arg\max_i \left\{ e_i^{\rm T} A y(t) \right\}, \\
j_{t+1} & \in \arg\min_j \left\{ x(t)^{\rm T} A e_j \right\}.
\end{aligned}
\end{equation}
The procedure may be viewed as a natural way through which two players could interact in a repeated game with stage game $(A,-A)$. The question is whether the sequence $(x(t), y(t))_t$ converges to something meaningful. 

In an elegant paper shortly after Brown's, Robinson showed that the average payoffs of the players in fictitious play converge to the value of the game~\cite{robinson1951iterative}. In particular, it was shown that
\begin{equation*}
\begin{aligned}
f_A(x(t), y(t)) & =\max_i e_i^{\rm T} A y(t) - \min_j x(t) A e_j  \\
              & \rightarrow 0, \text{ as }t\rightarrow \infty.
\end{aligned}
\end{equation*}
Hence, because $\min_j x(t) A e_j \le x(t)^{\rm T} A y(t) \le \max_i e_i^{\rm T} A y(t)$ and $\min_j x(t) A e_j \le z \le \max_i e_i^{\rm T} A y(t)$, it follows that all three quantities converge to the value of the game $z$.

Robinson's proof is an elegant induction argument, which eliminates one row or one column of $A$ at a time. Unraveling the induction, one can also deduce the following bound on the convergence rate of the procedure:
$$f_A(x(t), y(t)) = O(t^{-{1 \over m+n -2}}),$$
which appears rather slow, compared to the convergence rate of $O(t^{-\frac{1}{2}})$ that is typically achieved by no-regret learning algorithms~\cite{freund1999adaptive,littlestone1994weighted,cesa2006prediction}, and the improved convergence rate of $O({\log t \over t})$ of some no-regret learning algorithms, obtained recently~\cite{daskalakis2014near,RakhlinS13}. Indeed, about ten years after Robinson's proof and five decades ago, Samuel Karlin conjectured that the convergence rate of fictitious play should be $O(t^{-\frac{1}{2}})$, namely
\begin{conjecture}[{\cite{Karlin59}}] \label{Karlin's conjecture}
Fictitious play converges at rate $O(t^{-\frac{1}{2}})$ in all games.
\end{conjecture}

There is some evidence supporting a convergence rate of $O(t^{-\frac{1}{2}})$. As pointed out earlier, a convergence rate of $O(t^{-\frac{1}{2}})$ is quite common with dynamics that are known to converge. Indeed, a close relative of fictitious play, {\em follow the perturbed leader}, is known to achieve convergence rate of $O(t^{-\frac{1}{2}})$~\cite{cesa2006prediction}. Also, a continuous time version of fictitious play has been shown to converge in time $O(t^{-1})$~\cite{harris1998rate}. Despite this evidence and the apparent simplicity of fictitious play, the convergence rate from Robinson's proof has remained the state-of-the-art. Our main result is a counter-example, disproving Karlin's conjecture. If $I_n$ is the $n \times n$ identity matrix, we show the following:
\begin{theorem}\label{thm:main}
For every $n \ge 2$, fictitious play for $I_n$ may converge at rate $\Theta(t^{-\frac{1}{n}})$, if ties are broken arbitrarily.
\end{theorem}
Our counter-example, provided in Section~\ref{sec-counterexample}, constructs a valid execution of fictitious play for $I_n$ such that the empirical mixed strategies $x(t), y(t)$ of players satisfy
\begin{equation*}
\begin{aligned}
f_{I_n}(x(t), y(t))& = \max_i e_i^{\rm T} y(t) - \min_j x(t) e_j \\
                 & = \Theta(t^{-\frac{1}{n}}).
\end{aligned}
\end{equation*}

\begin{remark} It is crucial for our construction that ties in choosing a best response in~\eqref{eq:ties} can be broken arbitrarily at each step. This is allowed in Karlin's formulation of the conjecture. To distinguish this case from when ties are broken in some consistent way or randomly, we will call Karlin's conjecture with arbitrary tie-breaking  {\em Karlin's strong conjecture}, while that with lexicographic or random tie-breaking  {\em Karlin's weak conjecture}. With this terminology, Theorem~\ref{thm:main} disproves Karlin's strong conjecture.\end{remark}

Interestingly, like Robinson's upper bound argument, our lower bound also works by induction. We show that slow fictitious play executions for $I_2$ can be folded inside fictitious play executions for $I_3$, etc, leading to an exponentially slow convergence rate for fictitious play in $I_n$. More intuition about the construction is provided in Section~\ref{sec-preliminary}, and the complete details can be found in Section~\ref{sec-counterexample}.

While outperformed by modern learning algorithms~\cite{cesa2006prediction}, because of its simplicity, fictitious play was thought to provide a convincing explanation of Nash equilibrium play in zero-sum games. According to Luce and Raiffa ``Brown's results are not only computationally valuable but also quite illuminating from a substantive point of view. Imagine a pair of players repeating a game over and over again. It is plausible that at every stage a player attempts to exploit his knowledge of his opponent's past moves. Even though the game may be too complicated or too nebulous to be subjected to an adequate analysis, experience in repeated plays may tend to a statistical equilibrium whose (time) average return is approximately equal to the value of the game''~\cite{luce1957games}. In this light, our counterexample sheds doubt on the plausibility of fictitious play in explaining Nash equilibrium behavior. Given our counterexample, it is important to investigate whether fictitious play in random payoff zero-sum games satisfies Karlin's conjecture, or whether some choice of tie-breaking rule in the definition of fictitious play makes it satisfy Karlin's conjecture for all zero-sum games.  We did perform preliminary simulations of fictitious play with random tie-breaking on our lower bounding instances, as well as on  zero-sum games with i.i.d. uniform $[0,1]$ entries, and they suggest a quadratic rate of convergence. We leave a rigorous study of these important questions for future work.

\paragraph{Related Work} Fictitious play is one of the most well-studied dynamics in Game Theory, and we cannot do it justice in a short exposition. We only mention a few highlights here. As we have already mentioned, it was proposed by Brown, in a technical report at RAND corporation~\cite{Brown49}, and was shown to converge in two-person zero-sum games by Robinson~\cite{robinson1951iterative}. Miyakawa extended Robinson's results to  two-player games with two strategies per player assuming a specific tie-breaking rule~\cite{miyasawa1961convergence}, while Shapley constructed a two-player three-strategy game where fictitious play does not converge~\cite{shapley1964some}. Since then a lot of research has been devoted to understanding classes of games where fictitious play converges (e.g.~\cite{milgrom1991adaptive,monderer1996fictitious,hon1998learning,hahn1999convergence,sela2000fictitious,berger2005fictitious}) or does not converge~(e.g.~\cite{jordan1993three,gaunersdorfer1995fictitious,monderer19962,foster1998nonconvergence,krishna1998convergence}). Surveys can be found in~\cite{krishna1997learning,fudenberg1998theory,hofbauer2003evolutionary}. Other work has studied the approximation performance of fictitious play when used as a heuristic to find approximate Nash equilibria~\cite{conitzer2009approximation,goldberg2013approximation}.

In two-person zero-sum games, a convergence rate of $O(t^{-\frac{1}{m+n-2}})$ is implied by Robinson's proof, and S. Karlin conjectured that the convergence rate should be~$O(t^{-\frac{1}{2}})$, which would match what we know is achievable by no-regret learning algorithms~\cite{cesa2006prediction}. Indeed, Harris showed that a continuous analog of fictitious play converges in time $O(t^{-1})$~\cite{harris1998rate}. 
On the other hand, it is shown in~\cite{brandt2010rate} that it may take an exponential number of steps (in the size of the representation of the game) before any Nash equilibrium action is played by the players in fictitious play. However, this is not incompatible with Karlin's conjecture, since the payoffs may nevertheless still converge at rate $O(t^{-\frac{1}{2}})$. In fact, it is not even prohibited by~\cite{brandt2010rate} that the empirical strategies converge to Nash equilibrium strategies at rate $O(t^{-\frac{1}{2}})$.

As fictitious play is one of the simplest and most natural dynamics for learning in games it is widely used in applications, and has inspired several algorithms for learning and optimization, including von Neumann's variant of fictitious play for linear programming~\cite{neumann1954numerical}, the regret minimization paradigm~\cite{hannan1957approximation}, and lots of specialized algorithms in AI. See~\cite{brandt2010rate} for a survey.

\section{Preliminaries}\label{sec-preliminary}

\noindent {\bf Basic Definitions:} A \textit{two-player zero-sum game} can be represented by an $m \times n$ payoff matrix $A = (a_{ij})$, where $m$ and $n$ are the numbers of \textit{pure strategies} for the \textit{row player} and the \textit{column player}, respectively. The game is played when, simultaneously, the row player chooses one of his $m$ strategies, and the column player chooses one of her $n$ strategies. If the row player chooses strategy $i$ and the column player chooses strategy $j$, then the row player receives $a_{ij}$ from the column player.
  
The players can randomize their choices of strategies. A \textit{mixed strategy} for the row player is an $m$-vector $x$, where $x_i \geq 0$ and $\sum_i x_i = 1$. Similarly, a mixed strategy for the column player is an $n$-vector $y$, where $y_j \geq 0$ and $\sum_j y_j = 1$. When the players adopt those mixed strategies, the row player receives $x^{\rm T}Ay = \sum_{ij} a_{ij}x_iy_j$ in expectation from the column player.

A {\em min-max equilibrium}, or {\em Nash equilibrium}, of a zero-sum game $A$ is a pair of mixed strategies $\hat{x}$ for the row player and $\hat{y}$ for the column player such that Eq~\eqref{eq: minmax equilibrium} is satisfied. 

\medskip \noindent {\bf Dynamic:} We already described fictitious play in Section~\ref{sec-introduction}. We now introduce the notion of a dynamic as a formal way to describe a valid execution of fictitious play.
  
  For a vector $v$, let $\min v$ and $\max v$ denote its minimal and maximal components. A dynamic as defined in the next paragraph is a special case of a vector system as defined in \cite{robinson1951iterative} that starts from the zero vectors.
  
  \begin{definition}\label{system}
  A \textit{dynamic} $(U,V)$ for $A$ is a sequence of $n$-dimensional row vectors $U(0), \, U(1), \, \ldots$ and a sequence of $m$-dimensional column vectors $V(0), \, V(1), \, \ldots$ such that\footnote{Any vector presented using rectangular brackets is a column vector by default, unless it is followed by a transpose sign $\rm T$.}
\begin{equation*}
\begin{aligned}
U(0) & = [0,\ 0,\ \ldots,\ 0]^{\rm T}, \\
V(0) & = [0,\ 0,\ \ldots,\ 0],
\end{aligned}
\end{equation*} and
\begin{equation*}
\begin{aligned}
U(t+1) & = U(t) + e_i^{\rm T} A, \\
V(t+1) & = V(t) + A e_j,
\end{aligned}
\end{equation*} where $i$ and $j$ satisfy the conditions 
\begin{equation*}
\begin{aligned}
V_i(t) & = \max V(t), \\
U_j(t) & = \min U(t).
\end{aligned}
\end{equation*} 
  \end{definition}
  
  Just like there can be multiple valid executions of fictitious play for a matrix $A$, due to tie-breakings, there can be multiple possible dynamics for $A$. In fact, a dynamic for $A$ corresponds uniquely to an execution of fictitious play for $A$, if we identify $U(t)$ and $V(t)$ with $tx(t)^{\rm T} A$ and $tA y(t)$, respectively. (Recall from Section~\ref{sec-introduction} that $x(t)$ and $y(t)$ are the empirical mixed strategies of the two players for the first $t$ steps.)
  
  In terms of dynamics, Robinson's argument~\cite{robinson1951iterative} implies the following: If $(U,V)$ is a dynamic for an $m$ by $n$ matrix $A$, then \begin{equation*} \frac{\max V(t) - \min U(t)}{t} = O(t^{-\frac{1}{m+n-2}}). \end{equation*} Karlin's conjecture~\cite{Karlin59} amounts to the following: If $(U,V)$ is a dynamic for a matrix $A$, then \begin{equation*} \frac{\max V(t) - \min U(t)}{t} = O(t^{-\frac{1}{2}}). \end{equation*} Notice that in both equations above, the constant in $O(\cdot)$ may depend on $A$. Lastly, our construction implies that there exists a dynamic $(U,V)$ for $I_n$ such that \begin{equation*} \frac{\max V(t) - \min U(t)}{t} = \Theta(t^{-\frac{1}{n}}),\end{equation*} where the constant in $O(\cdot)$ may depend on $n$.
  
\smallskip \noindent {\bf Outline of our Construction:} First notice that, by Definition~\ref{system}, a dynamic $(U,V)$ for $I_n$ satisfies
\begin{equation*}
\begin{aligned}
U(0) & = [0, \, 0, \, \ldots, \, 0]^{\rm T}, \\
V(0) & = [0, \, 0, \, \ldots, \, 0],
\end{aligned}
\end{equation*} and 
\begin{equation*}
\begin{aligned}
U(t+1) & = U(t) + e_i^{\rm T}, \\
V(t+1) & = V(t) + e_j,
\end{aligned}
\end{equation*} where $i$ and $j$ satisfy the conditions
\begin{equation*}
\begin{aligned}
V_i(t) & = \max V(t), \\
U_j(t) & = \min U(t).
\end{aligned}
\end{equation*}
  
  A special property of the dynamics for $I_n$ is that permuting the $n$ components of every vector in a dynamic for $I_n$ by a common permutation $\sigma$ results in another dynamic for $I_n$, because $I_n$ stays the same when its rows and columns are both permuted by $\sigma$. This property allows us to combine many distinct cases in our main proof.
  
  For $n=2$, we can directly construct a dynamic for $I_2$ that converges at rate $\Theta(t^{-\frac{1}{2}})$, which we call the {\em main dynamic for $I_2$} (Figure~\ref{fig:main dynamic} and Claim~\ref{claim:main dynamic for n=2 claim}). At each step $t$, ties are simply broken by selecting the strategy that maximizes the ensuing gap $\max V(t) - \min U(t)$.
  
  For $n=3$, there is no obvious way to directly construct a dynamic for $I_3$ that converges at rate $\Theta(t^{-\frac{1}{3}})$. But, in the first three steps, it is easy to arrive at 
  \begin{equation*}
  \begin{aligned}
  U(3) & = [1, \, 1, \, 1]^{\rm T}, \\
  V(3) & = [0, \, 1, \, 2].
  \end{aligned}
  \end{equation*}
  Aiming for an inductive construction, let's in fact assume that, for some $P$, we can arrive at 
  \begin{equation*}
  \begin{aligned}
  U(3P) & = [P, \, P, \, P]^{\rm T}, \\
  V(3P) & = [Q_1, \, Q_2, \, Q_3],
  \end{aligned}
  \end{equation*} where $Q_1 \leq Q_2 \leq Q_3$. For the next few steps, we let $U$ increase only in its third component, and $V$ only in its first two components. We can do this as long as the third component of $V$, i.e. $Q_3$, remains its largest. Thus, we get to 
  \begin{align*}
  &U(3P+R) = [P, \, P, \, P+R]^{\rm T},\\ 
  &V(3P+R) = [Q_3, \, Q_3, \, Q_3].
  \end{align*}
  
The crucial component of our construction are the next steps, where we let $U$ and $V$ increase only their first two components, simulating a dynamic for the $2 \times 2$ subgame induced by the first two strategies of both players, i.e. $I_2$. (We are able to do this as long as the third component of $U$, i.e. $P+R$, remains its largest.) Since $U$ and $V$ have equal first and second components at step $3P+R$, any initial portion of any dynamic $(U',V')$ for $I_2$ can be copied, as long as the components of $U'$ remain at most $R$. Indeed, if we do this, then for all $t$ the first two components of $U(3P+R+t)$ are $P$ plus, respectively, the two components of $U'(t)$, and the first two components of $V(3P+R+t)$ are $Q_3$ plus, respectively, the two components of $V'(t)$. 
  
  For a dynamic $(U',V')$ for $I_2$, suppose that both components of $U'(t)$ are at most $R$, for all $t \le t_0$, for some~$t_0$. It can be easily checked that, if we copy this dynamic in the first two components of our dynamic $(U,V)$ for $I_3$ for $t_0$ steps, then the amount by which the gap for $(U,V)$ increases, that is, from \begin{equation*}\max V(3P+R) - \min U(3P+R)\end{equation*} to \begin{equation*}\max V(3P+R+t_0) - \min U(3P+R+t_0),\end{equation*} is exactly the gap $\max V'(t_0) - \min U'(t_0)$ of $(U',V')$ at $t_0$.
  
  We have two goals now. The first is to increase the gap for $(U,V)$ as much as possible, and the second is to come back to the pattern we started from (that is, $U$ has three equal components) so that we can apply the process again. To achieve our first goal, we want the gap $\max V'(t_0) - \min U'(t_0)$ to be as large as possible, subject to $\max U'(t_0) \leq R$. Naturally, we want $(U',V')$ to be the main dynamic for $I_2$, discussed earlier, as this achieves a rate of convergence of $\Theta(t^{-\frac{1}{2}})$. To achieve our second goal, we wish that $U'(t_0) = [R,R]^{\rm T}$, so that $U(3P+R+t_0) = [P+R,P+R,P+R]^{\rm T}$. Clearly, we must have $t_0 = 2R$ in this case. So, is it true that $U'(2R) = [R,R]^{\rm T}$, if $(U',V')$ is the main dynamic for $I_2$?
  
  From (Figure~\ref{fig:main dynamic}/Claim~\ref{claim:main dynamic for n=2 claim}), we see that there are indeed infinitely many $T$'s such that $U'(2T) = [T,T]^{\rm T}$. However, this is not true for all $T$. Thus, we can't exactly take $(U',V')$ to be the main dynamic for $I_2$, but will need a padded version of it. Hence, we define the {\em padding dynamic for $I_2$} as in Figure~\ref{fig:padding dynamics for n=2}/Claim~\ref{claim:main padding claim for n=2}, which reaches
  \begin{equation*}
  \begin{aligned}
  U''(2k) & = [k, \, k]^{\rm T}, \\
  V''(2k) & = [k-1, \, k+1],
  \end{aligned}
  \end{equation*} for all $k$. The dynamic $(U',V')$ that we copy into $(U,V)$ first follows the padding dynamic for $I_2$, and then the main dynamic for $I_2$. By picking the appropriate moment of transition, we can ensure that $(U',V')$ still converges at rate $\Theta(t^{-\frac{1}{2}})$, and $U'(2R) = [R,R]^{\rm T}$.
  
  Calculation shows that, if we repeat the process successively, the dynamic that will be obtained for $I_3$ converges at rate $\Theta(t^{-\frac{1}{3}})$. We call the resulting dynamic the {\em main dynamic for $I_3$}, and deal with $n=4$ in similar fashion, etc, leading to our main theorem.

\section{The Counterexample}\label{sec-counterexample}

In this section, we disprove Karlin's conjecture, by establishing the following.
  
  \begin{theorem}\label{example}
  For every $n \geq 2$, there exists a dynamic for $I_n$ such that for infinitely many $T$'s, \begin{equation*} \max V(nT) - \min U(nT) = \Theta(T^{\frac{n-1}{n}}). \end{equation*} 
  \end{theorem}
  \begin{prevproof}{Theorem}{example} Theorem~\ref{example} follows directly from Part 1 of the following Lemma (Part 2 is useful for showing Part 1 by induction):
  
  \begin{lemma}\label{key}  
  {\em Part 1:} For every $n \geq 2$, there exists a dynamic for $I_n$ such that for infinitely many $T$'s, \begin{equation*} U(nT) = [T, \, T, \, \ldots, T]^{\rm T}, \end{equation*} and \begin{equation*} \max V(nT) - \min U(nT) = \Theta(T^{\frac{n-1}{n}}). \end{equation*}
  
  {\em Part 2:} For every $n \geq 2$ and $T \geq 1$, there exists a dynamic for $I_n$ such that \begin{equation*} U(nT) = [T, \, T, \, \ldots, T]^{\rm T}, \end{equation*} and \begin{equation*} \max V(nT) - \min U(nT) = \Theta(T^{\frac{n-1}{n}}). \end{equation*} 
  
  In either part, the constant hidden by $\Theta(\cdot)$ may depend on $n$, but not on $T$.
  \end{lemma}
  
  \begin{prevproof}{Lemma}{key}
  We prove the lemma by induction on $n$. For each $n$, we prove Part 1 before Part 2.

\medskip \noindent {\bf Base case $n=2$:} We consider two dynamics for $I_2$, which we call the {\em padding dynamics}. The first steps of the padding dynamics are illustrated on the left and on the right respectively of Figure~\ref{fig:padding dynamics for n=2}. Notice that the strategy chosen by the row (respectively column) player at each step is exactly the index of the incremented component in $U$ (respectively $V$).

\begin{figure*}
\begin{equation*}\boxed{\begin{aligned}
 U(0) = [0,0]^{\rm T}, \quad & V(0) = [0,0] \quad&\quad U(0) = [0,0]^{\rm T}, \quad & V(0) = [0,0] \\
 \text{Step $1$:}~\text{row chooses $1$}\quad & \text{column chooses $2$} \quad&\quad \text{row chooses $1$}\quad &\text{column chooses $1$}\\
 U(1) = [1,0]^{\rm T}, \quad & V(1) = [0,1] \quad&\quad U(1) = [1,0]^{\rm T}, \quad & V(1) = [1,0] \\
 \text{Step $2$:}~\text{row chooses $2$}\quad & \text{column chooses $2$} \quad& \quad \text{row chooses $1$}\quad &\text{column chooses $2$}\\
 U(2) = [1,1]^{\rm T}, \quad & V(2) = [0,2] \quad&\quad U(2) = [2,0]^{\rm T}, \quad & V(2) = [1,1] \\
 \text{Step $3$:}~\text{row chooses $2$}\quad & \text{column chooses $1$} & \quad \text{row chooses $2$}\quad &\text{column chooses $2$}\\
 U(3) = [1,2]^{\rm T}, \quad & V(3) = [1,2] \quad&\quad U(3) = [2,1]^{\rm T}, \quad & V(3) = [1,2] \\
 \text{Step $4$:}~\text{row chooses $2$}\quad & \text{column chooses $1$} &\quad \text{row chooses $2$}\quad &\text{column chooses $2$}\\
 U(4) = [1,3]^{\rm T}, \quad & V(4) = [2,2] \quad&\quad U(4) = [2,2]^{\rm T}, \quad & V(4) = [1,3] \\
 \text{Step $5$:}~\text{row chooses $1$}\quad & \text{column chooses $1$} &\quad \text{row chooses $2$}\quad &\text{column chooses $1$}\\
 U(5) = [2,3]^{\rm T}, \quad & V(5) = [3,2] \quad&\quad U(5) = [2,3]^{\rm T}, \quad & V(5) = [2,3] \\
 \text{Step $6$:}~\text{row chooses $1$}\quad & \text{column chooses $1$} & \quad \text{row chooses $2$}\quad &\text{column chooses $1$}\\
 U(6) = [3,3]^{\rm T}, \quad & V(6) = [4,2] \quad&\quad U(6) = [2,4]^{\rm T}, \quad & V(6) = [3,3] \\
  \ldots, \quad & \ldots \quad & \quad \ldots, \quad & \ldots
  \end{aligned}} \end{equation*} \caption{The padding dynamics for $I_2$.}\label{fig:padding dynamics for n=2}
\end{figure*}

We claim the following.\begin{claim}\label{claim:padding dynamics for I2}
The dynamics shown in Figure~\ref{fig:padding dynamics for n=2} can be extended so that the dynamic on the left satisfies
\begin{equation}\label{eq:padding1}
\begin{aligned}
U(2k) & = [k, \, k]^{\rm T}, \\
V(2k) & = [k \pm 1, \, k \mp 1], 
\end{aligned}
\end{equation} 
for odd $k \geq 1$, while the dynamic on the right satisfies \eqref{eq:padding1} for even $k \geq 2$. The choice of $+$ or $-$ depends on the parity of $\lceil \frac{k}{2} \rceil$.
\end{claim}
\begin{prevproof}{Claim}{claim:padding dynamics for I2}
To see the claim for the dynamic on the left, compare $U(t),V(t)$ at steps $t=2$ and $t=6$. The two components of $U(t)$ are equal, while the two components of $V(t)$ differ by $2$. So, after exchanging the strategies $1 \leftrightarrow 2$, we can repeat the players' choices at Steps $3$, $4$, $5$ and $6$ in Steps $7, 8, 9$ and $10$ respectively to arrive at $U(10)=[5,5]^{\rm T}$ and $V(10)=[4,6]$. And, we can continue the same way ad infinitum, which proves the claim for all odd $k$'s. Similar argument for the dynamic on the right proves for all even $k$'s.
\end{prevproof}    
 
 By using either of the padding dynamics for $I_2$ and exchanging the components as necessary, we see the following:
 \begin{claim}\label{claim:main padding claim for n=2}
 For any $k \geq 1$, there exists a padding dynamic for $I_2$ such that
 \begin{equation*}\label{pad2}
 \begin{aligned}
 U(2k) & = [k, \, k]^{\rm T}, \\
 V(2k) & = [k - 1, \, k + 1].
 \end{aligned}
 \end{equation*}
  \end{claim}
  Next, we define the \textit{main dynamic for $I_2$}, whose first steps are shown in Figure~\ref{fig:main dynamic} in the appendix. We claim the following.
  \begin{claim} \label{claim:main dynamic for n=2 claim}
The dynamic given in Figure~\ref{fig:main dynamic} can be extended so that it satisfies the following for all $k \ge 1$:
\begin{equation}\label{gapbound2}
\begin{aligned}
    & U(2k(2k-1)) \\
    & \quad \quad = [k(2k-1), \, k(2k-1)]^{\rm T}, \\
    & V(2k(2k-1)) \\
    & \quad \quad = [(k \pm 1)(2k-1), \, (k \mp 1)(2k-1)],
\end{aligned}
\end{equation} where the choice of $+$ or $-$ depends on the parity of $k$. \label{claim:main dynamic for n=2}
  \end{claim}
  \begin{prevproof}{Claim}{claim:main dynamic for n=2} This can be easily established by induction on $k$. Indeed, Figure~\ref{fig:main dynamic} establishes the claim for $k=1, 2, 3$. In general, suppose that, for some $k$:
  \begin{align*}
  & U(2k(2k-1)) \\ 
  & \quad \quad = [k(2k-1), \, k(2k-1)]^{\rm T}, \\
  & V(2k(2k-1)) \\
  & \quad \quad = [(k + 1)(2k-1), \, (k - 1)(2k-1)].
  \end{align*}
 
  Generalizing what is taking place from Step $13$ through Step $30$ of Figure~\ref{fig:main dynamic}, the dynamic proceeds with both players playing strategy $1$ for one step, the row player playing strategy $1$ and the column player playing strategy $2$ for the next $4k$ steps, and both players playing strategy $2$ for the next $4k+1$ steps, resulting in
  \begin{align*}
  & U(2(k+1)(2(k+1)-1)) \\
  & \quad = [(k+1)(2(k+1)-1), \, (k+1)(2(k+1)-1)]^{\rm T},\\
  & V(2(k+1)(2(k+1)-1)) \\
  & \quad = [k(2(k+1)-1), \, (k+2)(2(k+1)-1)].
  \end{align*}
This establishes the claim for $k+1$. The derivation is similar, if for $k$ Equation~\eqref{gapbound2} is satisfied with $\pm$ and $\mp$ instantiated by $-$ and $+$ respectively.
  \end{prevproof}
  
  Notice that Claim~\ref{claim:main dynamic for n=2} proves Part 1 of Lemma~\ref{key} for $n=2$.
       
  Now, for any given $T$, we construct a dynamic $(U',V')$ for $I_2$ that satisfies the conditions in Part 2 of Lemma~\ref{key}. Let $k$ be the largest integer such that $k(2k-1) \leq T$, and $l= T-k(2k-1)+1$. Starting with $U'(0) = [0,0]^{\rm T}$ and $V'(0) = [0,0]$, we first evolve the vectors to
  \begin{equation*}
  \begin{aligned}
  U'(2l) & = [l, \, l]^{\rm T}, \\
  V'(2l) & = [l - 1, \, l + 1],
  \end{aligned}
  \end{equation*} as enabled by Claim~\ref{claim:main padding claim for n=2}. Because the components of $U'(2l)$ and $V'(2l)$ are exactly $l-1$ larger than the corresponding components of $U(2)$ and $V(2)$ of the main dynamic for $I_2$, we can further evolve the vectors $U'$ and $V'$ for $2k(2k-1)-2$ steps, mirroring the players' choices from Steps $3$ through $2k(2k-1)$ in the main dynamic for $I_2$. Using Claim~\ref{claim:main dynamic for n=2}, we arrive at
  \begin{equation*}
  \begin{aligned}
  U'(2T) & = [T, \, T]^{\rm T}, \\
  V'(2T) & = [T \pm (2k-1), \, T \mp (2k-1)],
  \end{aligned}
  \end{equation*} which satisfies 
  \begin{equation*}
  \begin{aligned}
  & \quad \max V'(2T) - \min U'(2T) \\
  & = 2k-1 \\
  & = \Theta(T^{\frac{1}{2}}). 
  \end{aligned}
  \end{equation*} The constant hidden by $\Theta(\cdot)$ can obviously be chosen uniformly for all $T$. We have thus proved Part 2 of Lemma~\ref{key} for $n = 2$.
    
\medskip \noindent {\bf Induction Step:} Assume that Lemma~\ref{key} is true for a certain $n \geq 2$. To prove it for $n+1$, we first consider two {\em padding dynamics for $I_{n+1}$}, whose first steps are shown in Figure~\ref{fig:padding dynamics for general n} (in the appendix). We suppress the step numbers and strategy choices in the figure, since these can be easily inferred from the vectors. These dynamics generalize the padding dynamics for $I_2$ appropriately.
  Similarly to Claim~\ref{claim:main padding claim for n=2}, we can show the following:
  \begin{claim}\label{claim:padding dynamics for In}
  For any $k \geq 1$, there exists a padding dynamic for $I_{n+1}$ such that
  \begin{equation*}
  \begin{aligned}
  U((n+1)k) & = [k, \, k, \, \ldots , \, k]^{\rm T}, \\
  V((n+1)k) & = [k - 1, \, k, \, \ldots , \, k, \, k + 1]. 
  \end{aligned}
  \end{equation*}
  \end{claim}
  \begin{prevproof}{Claim}{claim:padding dynamics for In}
We omit most of the details as the proof is very similar to that of Claim~\ref{claim:main padding claim for n=2}. For example, in the top dynamic in Figure~\ref{fig:padding dynamics for general n}, we see that $U$ reaches both $[1,1, \ldots, 1]^{\rm T}$ and $[3,3, \dots, 3]^{\rm T}$. Since the corresponding values for $V$ have the same format up to an additive shift and a permutation of the components, we can repeat the pattern ad infinitum to prove the cases for odd $k$'s. Similarly, the bottom dynamic in Figure~\ref{fig:padding dynamics for general n} deals with even $k$'s.
\end{prevproof}  
 
  \smallskip Next, we define the main dynamic for $I_{n+1}$, which pieces together parts of various dynamics for $I_n$ obtained from the inductive hypothesis. We describe this dynamic inductively by dividing it into \textit{epochs}:
  \begin{enumerate}
  \item  {\em Initial steps leading to 1st epoch:} Starting with $U(0) = [0,\ldots,0]^{\rm T}$ and $V(0) = [0,\ldots,0]$, we first evolve the vectors to
  \begin{equation}\label{firstepoch}
  \begin{aligned}
  U(n+1) & = [1, \, 1, \, \ldots , \, 1]^{\rm T}, \\
  V(n+1) & = [0, \, 1, \, \ldots , \, 1, \, 2],
  \end{aligned}
  \end{equation} as enabled by Claim~\ref{claim:padding dynamics for In}. We mark those vectors as the beginning of the $1$st epoch.
  
 \item {\em Evolution within an epoch:} For $i \ge 1$, suppose that at the beginning of the $i$-th epoch we satisfy 
 \begin{equation*}
 \begin{aligned}
 U((n+1)P) & = [P, \, P, \, \ldots , \, P]^{\rm T}, \\
 V((n+1)P) & = [Q_1, \, Q_2, \, \ldots , \, Q_{n+1}]. 
 \end{aligned}
 \end{equation*} Without loss of generality, let us also assume that $Q_1 \leq Q_2 \leq \cdots \leq Q_{n+1}$.
  
  Because $(n+1)P = \sum_j Q_j$, we have \begin{equation*} (n+1)(Q_{n+1} - P) = \sum_j (Q_{n+1} - Q_j). \end{equation*} For the next $R = (n+1)(Q_{n+1} - P)$ steps, let $U$ increase only in its $(n+1)$-th component, and $V$ increase $Q_{n+1} - Q_j$ times its $j$-th component, for all $j$ (the exact order of those increments doesn't matter). The process is compatible with the definition of a dynamic because, in each of those $R$ steps, the $(n+1)$-th component of $V$ remains maximal in $V$, and the first $n$ components of $U$ remain minimal in $U$. At the end of these steps, we arrive at 
  \begin{equation}\label{midepoch} 
  \begin{aligned}
  U((n+1)Q_{n+1}) & = [P, \, \ldots , \, P, \, P+R]^{\rm T}, \\
  V((n+1)Q_{n+1}) & = [Q_{n+1}, \, \ldots , \, Q_{n+1}].
  \end{aligned}
  \end{equation}
  
  Now, from our inductive hypothesis, there exists a dynamic $(\hat{U}, \hat{V})$ for $I_n$ such that 
  \begin{equation*}
  \begin{aligned}
  \hat{U}(nR) & = [R, \, R, \, \ldots , \, R]^{\rm T}, \\
  \hat{V}(nR) & = [S_1, \, S_2, \, \ldots , \, S_n],
  \end{aligned}
  \end{equation*} and 
  \begin{equation*}
  \max \hat{V}(nR) - \min \hat{U}(nR) = \Theta(R^{\frac{n-1}{n}}),
  \end{equation*}
  where the constant hidden by $\Theta(\cdot)$ is independent of $R$. Starting from (\ref{midepoch}), for the next $nR$ steps, we increment only the first $n$ components of $U$ and $V$, in a way that mirrors the strategy choices of the players in the evolution of $\hat{U}$ and $\hat{V}$, starting from $\hat{U}(0) = [0,\ldots,0]^{\rm T}$ and $\hat{V}(0) = [0,\ldots,0]$, until $\hat{U}(nR)$ and $\hat{V}(nR)$. Because the $(n+1)$-th component of $V$ remains minimal in $V$, we see that, in each of those $nR$ steps, a maximal component among the first $n$ components of $V$ is also a maximal component of the entire vector $V$. Similarly, a minimal component among the first $n$ components of $U$ is also a minimal component of the entire vector $U$. Therefore, the process is compatible with the definition of a dynamic. At the end of the $nR$ steps, we have
  \begin{equation*}
  \begin{aligned}
  & U((n+1)(P+R)) \\
  & \quad = [P+R, \, \ldots , \, P+R, \, P+R]^{\rm T}, \\
  & V((n+1)(P+R)) \\
  & \quad = [Q_{n+1}+S_1, \, \ldots , \, Q_{n+1}+S_n, \, Q_{n+1}], 
  \end{aligned}
  \end{equation*} which we mark as the beginning of the $(i+1)$-th epoch.
Notice that the vectors have a format that allows the induction to continue.
  \end{enumerate}

  We analyze the convergence rate of the main dynamic for $I_{n+1}$. For each $i$, let $(n+1)T_i$ be the step number at the beginning of the $i$-th epoch, and $G_i$ the gap
  \begin{equation*}
  G_i = \max V((n+1)T_i) - \min U((n+1)T_i).
  \end{equation*} Using the $P$, $Q$, $R$, and $S$ notation above, we have the following relations:
  \begin{equation*}
  \begin{aligned}
  T_i & = P, \\
  T_{i+1} & = P+R, \\
  G_i & = Q_{n+1} - P, \\
  G_{i+1} & = \max_j (S_j + Q_{n+1}) - (P + R) \\
          & = (Q_{n+1} - P) + (\max \hat{V}(nR) - \min \hat{U}(nR)) \\
          & = (Q_{n+1} - P) + \Theta(R^{\frac{n-1}{n}}), \\
  R & = (n+1)(Q_{n+1} - P).
  \end{aligned}
  \end{equation*}
  From the above, along with the initial values from (\ref{firstepoch}), we obtain the following recursive relations: \begin{equation*}
  \begin{aligned}
  G_1 &= 1, \\
  T_1 &= 1, \\
  G_{i+1} & = G_i + \Theta([(n+1)G_i]^{\frac{n-1}{n}}), \\
  T_{i+1} & = T_i + (n+1)G_i,
  \end{aligned}
  \end{equation*} where the constants hidden by the $\Theta(\cdot)$'s depend only on $n+1$. A simple calculation based on those relations yields 
  \begin{equation*}
  \begin{aligned}
  G_i & = \Theta(i^n), \\
  T_i & = \Theta(i^{n+1}), 
  \end{aligned}
  \end{equation*} and so
  \begin{equation*}
  G_i = \Theta(T_i^{\frac{n}{n+1}}),
  \end{equation*}
  where the constants hidden by the $\Theta(\cdot)$'s depend only on $n+1$. Consequently, by considering the beginning of each of the infinitely many epoches, the main dynamic for $I_{n+1}$ satisfies Part 1 of Lemma~\ref{key} for $n+1$.
  
  \medskip We are now ready to construct, for any given $T$, a dynamic $(U',V')$ for $I_{n+1}$ satisfying the conditions in Part 2. Let $k$ be the largest integer so that $T_k \leq T$, and $l= T-T_k+1$. Starting from $U'(0) =[0,\ldots,0]^{\rm T}$ and $V'(0) = [0,\ldots,0]$, we first evolve the vectors to 
  \begin{equation*} 
  \begin{aligned}
  U'((n+1)l) & = [l, \, l, \, \ldots , \, l]^{\rm T}, \\
  V'((n+1)l) & = [l-1, \, l, \, \ldots , \, l, \, l+1],
  \end{aligned}
  \end{equation*} as enabled by Claim~\ref{claim:padding dynamics for In}. Because the components of $U'((n+1)l)$ and $V'((n+1)l)$ are exactly $l-1$ larger than the corresponding components of $U(n+1)$ and $V(n+1)$ in the main dynamic for $I_{n+1}$ (i.e. the vectors marking the beginning of the $1$st epoch), we can further evolve the vectors $U'$ and $V'$ for $(n+1)T_k - (n+1)$ steps, mirroring the players' choices in Steps $n+2$ through $(n+1)T_k$ (i.e. up until the beginning of the $k$-th epoch) in the main dynamic for $I_{n+1}$. The components of $U'((n+1)T)$ and $V'((n+1)T)$ at the end of this process are $l-1$ plus the corresponding components of $U((n+1)T_k)$ and $V((n+1)T_k)$ in the main dynamic for $I_{n+1}$. Thus, we have \begin{equation*}
  U'((n+1)T) = [T, \, T, \, \ldots , \, T]^{\rm T},
  \end{equation*} and 
  \begin{equation*}
  \begin{aligned}
    & \quad \max V'((n+1)T) - \min U'((n+1)T) \\
    & = G_k \\
    & = \Theta(T_k^{\frac{n}{n+1}}) \\
    & = \Theta(T^{\frac{n}{n+1}}). 
  \end{aligned}
  \end{equation*} The constant hidden by the $\Theta(\cdot)$'s can obviously be chosen uniformly for all $T$. We have thus proved Part 2 of Lemma~\ref{key} for $n+1$. By induction, the proof of Lemma~\ref{key} is completed.
  \end{prevproof} 
  \end{prevproof}

\begin{remark}
Notice that, even though we do not explicitly state it in Theorem~\ref{example}, our proof implies something stronger, namely that for every $n \geq 2$, there exists a dynamic for $I_n$ such that for all $t$ (as opposed to just infinitely many $t$'s):
\begin{equation*} \max V(t) - \min U(t) = \Theta(t^{\frac{n-1}{n}}). \end{equation*} 
\end{remark}
\bibliographystyle{abbrv}
\bibliography{costasbibold} 

\begin{thebibliography}{10}

\bibitem{Adler13}
I.~Adler.
\newblock The equivalence of linear programs and zero-sum games.
\newblock {\em International Journal of Game Theory}, 42(1):165--177, 2013.

\bibitem{Interview_with_Dantzig}
D.~J. Albers and C.~Reid.
\newblock An interview with {G}eorge {B.} {D}antzig: the father of linear
  programming.
\newblock {\em The College Mathematics Journal}, 17(4):293--314, 1986.

\bibitem{aumann1987game}
R.~Aumann.
\newblock Game theory.
\newblock {\em The new palgrave---a dictionary of Economics. Edited by John
  Eatwell, Murray Milgate and Peter Newman. The Macmillan Press Limited}, 1987.

\bibitem{berger2005fictitious}
U.~Berger.
\newblock Fictitious play in 2$\times$n games.
\newblock {\em Journal of Economic Theory}, 120(2):139--154, 2005.

\bibitem{brandt2010rate}
F.~Brandt, F.~Fischer, and P.~Harrenstein.
\newblock On the rate of convergence of fictitious play.
\newblock {\em Theory of Computing Systems}, 53:41--52, 2013.

\bibitem{Brown49}
G.~W. Brown.
\newblock Some notes on computation of games solutions.
\newblock Technical report, P-78, The Rand Corporation, 1949.

\bibitem{brown1951iterative}
G.~W. Brown.
\newblock Iterative solution of games by fictitious play.
\newblock {\em Activity Analysis of Production and Allocation}, 13(1):374--376,
  1951.

\bibitem{cesa2006prediction}
N.~Cesa-Bianchi and G.~Lugosi.
\newblock {\em Prediction, learning, and games}.
\newblock Cambridge University Press, 2006.

\bibitem{conitzer2009approximation}
V.~Conitzer.
\newblock Approximation guarantees for fictitious play.
\newblock In {\em the 47th Annual Allerton Conference on Communication,
  Control, and Computing}, pages 636--643. IEEE, 2009.

\bibitem{Dantzig63}
G.~B. Dantzig.
\newblock {\em Linear {P}rogramming and {E}xtensions}.
\newblock Princeton University Press, 1963.

\bibitem{daskalakis2014near}
C.~Daskalakis, A.~Deckelbaum, and A.~Kim.
\newblock Near-optimal no-regret algorithms for zero-sum games.
\newblock {\em Games and Economic Behavior}, 2014.

\bibitem{foster1998nonconvergence}
D.~P. Foster and H.~P. Young.
\newblock On the nonconvergence of fictitious play in coordination games.
\newblock {\em Games and Economic Behavior}, 25(1):79--96, 1998.

\bibitem{freund1999adaptive}
Y.~Freund and R.~E. Schapire.
\newblock Adaptive game playing using multiplicative weights.
\newblock {\em Games and Economic Behavior}, 29(1):79--103, 1999.

\bibitem{fudenberg1998theory}
D.~Fudenberg.
\newblock {\em The theory of learning in games}, volume~2.
\newblock MIT press, 1998.

\bibitem{gaunersdorfer1995fictitious}
A.~Gaunersdorfer and J.~Hofbauer.
\newblock Fictitious play, {S}hapley polygons, and the replicator equation.
\newblock {\em Games and Economic Behavior}, 11(2):279--303, 1995.

\bibitem{goldberg2013approximation}
P.~W. Goldberg, R.~Savani, T.~B. S{\o}rensen, and C.~Ventre.
\newblock On the approximation performance of fictitious play in finite games.
\newblock {\em International Journal of Game Theory}, 42(4):1059--1083, 2013.

\bibitem{hahn1999convergence}
S.~Hahn.
\newblock The convergence of fictitious play in 3$\times$3 games with strategic
  complementarities.
\newblock {\em Economics Letters}, 64(1):57--60, 1999.

\bibitem{hannan1957approximation}
J.~Hannan.
\newblock Approximation to {B}ayes risk in repeated play.
\newblock {\em Contributions to the Theory of Games}, 3:97--139, 1957.

\bibitem{harris1998rate}
C.~Harris.
\newblock On the rate of convergence of continuous-time fictitious play.
\newblock {\em Games and Economic Behavior}, 22(2):238--259, 1998.

\bibitem{hofbauer2003evolutionary}
J.~Hofbauer and K.~Sigmund.
\newblock Evolutionary game dynamics.
\newblock {\em Bulletin of the American Mathematical Society}, 40(4):479--519,
  2003.

\bibitem{hon1998learning}
S.~Hon-Snir, D.~Monderer, and A.~Sela.
\newblock A learning approach to auctions.
\newblock {\em Journal of Economic Theory}, 82(1):65--88, 1998.

\bibitem{jordan1993three}
J.~S. Jordan.
\newblock Three problems in learning mixed-strategy {N}ash equilibria.
\newblock {\em Games and Economic Behavior}, 5(3):368--386, 1993.

\bibitem{Karlin59}
S.~Karlin.
\newblock {\em Mathematical Methods and Theory in Games, Programming, and
  Economics}.
\newblock Addison-Wesley, 1959.

\bibitem{krishna1997learning}
V.~Krishna and T.~Sj{\"o}str{\"o}m.
\newblock Learning in games: Fictitious play dynamics.
\newblock In {\em Cooperation: Game-Theoretic Approaches}, pages 257--273.
  Springer, 1997.

\bibitem{krishna1998convergence}
V.~Krishna and T.~Sj{\"o}str{\"o}m.
\newblock On the convergence of fictitious play.
\newblock {\em Mathematics of Operations Research}, 23(2):479--511, 1998.

\bibitem{littlestone1994weighted}
N.~Littlestone and M.~K. Warmuth.
\newblock The weighted majority algorithm.
\newblock {\em Information and Computation}, 108(2):212--261, 1994.

\bibitem{luce1957games}
R.~D. Luce and H.~Raiffa.
\newblock {\em Games and Decisions: Introduction and Critical Survey}.
\newblock Courier Dover Publications, 1957.

\bibitem{milgrom1991adaptive}
P.~Milgrom and J.~Roberts.
\newblock Adaptive and sophisticated learning in normal form games.
\newblock {\em Games and Economic Behavior}, 3(1):82--100, 1991.

\bibitem{miyasawa1961convergence}
K.~Miyasawa.
\newblock On the convergence of the learning process in a 2 x 2 non-zero-sum
  two-person game.
\newblock Technical report, DTIC Document, 1961.

\bibitem{monderer19962}
D.~Monderer and A.~Sela.
\newblock A 2$\times$2 game without the fictitious play property.
\newblock {\em Games and Economic Behavior}, 14(1):144--148, 1996.

\bibitem{monderer1996fictitious}
D.~Monderer and L.~S. Shapley.
\newblock Fictitious play property for games with identical interests.
\newblock {\em Journal of Economic Theory}, 68(1):258--265, 1996.

\bibitem{RakhlinS13}
A.~Rakhlin and K.~Sridharan.
\newblock Optimization, learning, and games with predictable sequences.
\newblock In {\em the 27th Annual Conference on Neural Information Processing
  Systems (NIPS)}, 2013.

\bibitem{robinson1951iterative}
J.~Robinson.
\newblock An iterative method of solving a game.
\newblock {\em Annals of Mathematics}, pages 296--301, 1951.

\bibitem{sela2000fictitious}
A.~Sela.
\newblock Fictitious play in 2$\times$ 3 games.
\newblock {\em Games and Economic Behavior}, 31(1):152--162, 2000.

\bibitem{shapley1964some}
L.~S. Shapley.
\newblock Some topics in two-person games.
\newblock {\em Advances in Game Theory}, 52:1--29, 1964.

\bibitem{vN}
J.~von Neumann.
\newblock {Zur Theorie der Gesellshaftsspiele}.
\newblock {\em Mathematische Annalen}, 100:295--320, 1928.

\bibitem{neumann1954numerical}
J.~von Neumann.
\newblock A numerical method to determine optimum strategy.
\newblock {\em Naval Research Logistics Quarterly}, 1(2):109--115, 1954.

\end{thebibliography}
 

%

%

\newpage

  \begin{figure*}[h!]
\centering  APPENDIX\\
  \begin{equation*}\boxed{\label{maindynamic2} \begin{aligned}
~~~~~~~~~~~~~~~~~~ U(0) = [0,0]^{\rm T}, \quad & \quad V(0) = [0,0] ~~~~~~~~~~~~~~~~~~~~~\\
 \text{Step $1$:}~\text{row chooses $1$}\quad & \text{column chooses $2$} \\
 U(1) = [1,0]^{\rm T}, \quad & \quad V(1) = [0,1] \\
 \text{Step $2$:}~\text{row chooses $2$}\quad & \text{column chooses $2$} \\
 U(2) = [1,1]^{\rm T}, \quad & \quad V(2) = [0,2] \\
 \text{Step $3$:}~\text{row chooses $2$}\quad & \text{column chooses $2$} \\
 U(3) = [1,2]^{\rm T}, \quad & \quad V(3) = [0,3] \\
 \text{Step $4$:}~\text{row chooses $2$}\quad & \text{column chooses $1$} \\
 U(4) = [1,3]^{\rm T}, \quad & \quad V(4) = [1,3] \\
 \text{Step $5$:}~\text{row chooses $2$}\quad & \text{column chooses $1$} \\
 U(5) = [1,4]^{\rm T}, \quad & \quad V(5) = [2,3] \\
 \text{Step $6$:}~\text{row chooses $2$}\quad & \text{column chooses $1$} \\
 U(6) = [1,5]^{\rm T}, \quad & \quad V(6) = [3,3] \\
 \text{Step $7$:}~\text{row chooses $2$}\quad & \text{column chooses $1$} \\
 U(7) = [1,6]^{\rm T}, \quad & \quad V(7) = [4,3] \\
 \text{Step $8$:}~\text{row chooses $1$}\quad & \text{column chooses $1$} \\
 U(8) = [2,6]^{\rm T}, \quad & \quad V(8) = [5,3] \\
 \ldots, \quad & \quad \ldots \\
 \text{Step $12$:}~\text{row chooses $1$}\quad & \text{column chooses $1$} \\
 U(12) = [6,6]^{\rm T}, \quad & \quad V(12) = [9,3] \\
 \text{Step $13$:}~\text{row chooses $1$}\quad & \text{column chooses $1$} \\
 U(13) = [7,6]^{\rm T}, \quad & \quad V(13) = [10,3] \\
 \text{Step $14$:}~\text{row chooses $1$}\quad & \text{column chooses $2$} \\
 U(14) = [8,6]^{\rm T}, \quad & \quad V(14) = [10,4] \\
 \ldots, \quad & \quad \ldots \\
 \text{Step $20$:}~\text{row chooses $1$}\quad & \text{column chooses $2$} \\
 U(20) = [14,6]^{\rm T}, \quad & \quad V(20) = [10,10] \\
 \text{Step $21$:}~\text{row chooses $1$}\quad & \text{column chooses $2$} \\
 U(21) = [15,6]^{\rm T}, \quad & \quad V(21) = [10,11] \\
 \text{Step $22$:}~\text{row chooses $2$}\quad & \text{column chooses $2$} \\
 U(22) = [15,7]^{\rm T}, \quad & \quad V(22) = [10,12] \\
 \ldots, \quad & \quad \ldots \\
 \text{Step $30$:}~\text{row chooses $2$}\quad & \text{column chooses $2$} \\
 U(30) = [15,15]^{\rm T}, \quad & \quad V(30) = [10,20] \\
 \ldots, \quad & \quad \ldots
  \end{aligned}} \end{equation*}\caption{The main dynamic for $I_2$.} \label{fig:main dynamic}
  \end{figure*}
  
\begin{figure*}
  \begin{equation*} \boxed{\begin{aligned}
  U(0) = [0,\ 0,\ 0,\ \ldots ,\ 0]^{\rm T}, \quad & \quad V(0) = [0,\ 0,\ 0,\ \ldots ,\ 0] \\
  U(1) = [1,\ 0,\ 0,\ \ldots ,\ 0]^{\rm T}, \quad & \quad V(1) = [0,\ 1,\ 0,\ \ldots ,\ 0] \\
  U(2) = [1,\ 1,\ 0,\ \ldots ,\ 0]^{\rm T}, \quad & \quad V(2) = [0,\ 1,\ 1,\ \ldots ,\ 0] \\
  \ldots, \quad & \quad \ldots \\
  U(n) = [1,\ 1,\ 1,\ \ldots ,\ 1,\ 0]^{\rm T}, \quad & \quad V(n) = [0,\ 1,\ 1,\ \ldots ,\ 1,\ 1] \\
  U(n+1) = [1,\ 1,\ 1,\ \ldots ,\ 1,\ 1]^{\rm T}, \quad & \quad V(n+1) = [0,\ 1,\ 1,\ \ldots ,\ 1,\ 2] \\
  U(n+2) = [1,\ 1,\ 1,\ \ldots ,\ 1,\ 2]^{\rm T}, \quad & \quad V(n+2) = [1,\ 1,\ 1,\ \ldots ,\ 1,\ 2] \\
  U(n+3) = [1,\ 1,\ 1,\ \ldots ,\ 1,\ 3]^{\rm T}, \quad & \quad V(n+3) = [2,\ 1,\ 1,\ \ldots ,\ 1,\ 2] \\
  U(n+4) = [2,\ 1,\ 1, \ldots ,\ 1,\ 3]^{\rm T}, \quad & \quad V(n+4) = [2,\ 2,\ 1,\ \ldots ,\ 1,\ 2] \\
  \ldots, \quad & \quad \ldots \\
  U(2n+2) = [2,\ 2,\ 2,\ \ldots ,\ 2,\ 1,\ 3]^{\rm T}, \quad & \quad V(2n+2) = [2,\ 2,\ 2,\ \ldots ,\ 2,\ 2,\ 2] \\
  U(2n+3) = [2,\ 2,\ 2,\ \ldots ,\ 2,\ 2,\ 3]^{\rm T}, \quad & \quad V(2n+3) = [2,\ 2,\ 2,\ \ldots ,\ 2,\ 3,\ 2] \\
  U(2n+4) = [2,\ 2,\ 2,\ \ldots ,\ 2,\ 3,\ 3]^{\rm T}, \quad & \quad V(2n+4) = [3,\ 2,\ 2,\ \ldots ,\ 2,\ 3,\ 2] \\
  U(2n+5) = [3,\ 2,\ 2,\ \ldots ,\ 2,\ 3,\ 3]^{\rm T}, \quad & \quad V(2n+5) = [3,\ 3,\ 2,\ \ldots ,\ 2,\ 3,\ 2] \\
  \ldots, \quad & \quad \ldots \\
  U(3n+2) = [3,\ 3,\ 3,\ \ldots ,\ 3,\ 2,\ 3,\ 3]^{\rm T}, \quad & \quad V(3n+2) = [3,\ 3,\ 3,\ \ldots ,\ 3,\ 3,\ 3,\ 2] \\
  U(3n+3) = [3,\ 3,\ 3,\ \ldots ,\ 3,\ 3,\ 3,\ 3]^{\rm T}, \quad & \quad V(3n+3) = [3,\ 3,\ 3,\ \ldots ,\ 3,\ 4,\ 3,\ 2] \\
  \ldots, \quad & \quad \ldots,
  \end{aligned}} \end{equation*} \centering and
  \begin{equation*} \boxed{\begin{aligned}
  U(0) = [0,\ 0,\ 0,\ \ldots ,\ 0]^{\rm T}, \quad & \quad V(0) = [0,\ 0,\ 0,\ \ldots ,\ 0] \\
  U(1) = [1,\ 0,\ 0,\ \ldots ,\ 0]^{\rm T}, \quad & \quad V(1) = [1,\ 0,\ 0,\ \ldots ,\ 0] \\
  U(2) = [2,\ 0,\ 0,\ \ldots ,\ 0]^{\rm T}, \quad & \quad V(2) = [1,\ 1,\ 0,\ \ldots ,\ 0] \\
  U(3) = [2,\ 1,\ 0,\ \ldots ,\ 0]^{\rm T}, \quad & \quad V(3) = [1,\ 1,\ 1,\ \ldots ,\ 0] \\
  \ldots, \quad & \quad \ldots \\
  U(n+1) = [2,\ 1,\ 1,\ \ldots ,\ 1,\ 0]^{\rm T}, \quad & \quad V(n+1) = [1,\ 1,\ 1,\ \ldots ,\ 1,\ 1] \\
  U(n+2) = [2,\ 1,\ 1,\ \ldots ,\ 1,\ 1]^{\rm T}, \quad & \quad V(n+2) = [1,\ 1,\ 1,\ \ldots ,\ 1,\ 2] \\
  U(n+3) = [2,\ 1,\ 1,\ \ldots ,\ 1,\ 2]^{\rm T}, \quad & \quad V(n+3) = [1,\ 2,\ 1,\ \ldots ,\ 1,\ 2] \\
  U(n+4) = [2,\ 2,\ 1,\ \ldots ,\ 1,\ 2]^{\rm T}, \quad & \quad V(n+4) = [1,\ 2,\ 2,\ \ldots ,\ 1,\ 2] \\
  \ldots, \quad & \quad \ldots \\
  U(2n+1) = [2,\ 2,\ 2,\ \ldots ,\ 2,\ 1,\ 2]^{\rm T}, \quad & \quad V(2n+1) = [1,\ 2,\ 2,\ \ldots ,\ 2,\ 2] \\
  U(2n+2) = [2,\ 2,\ 2,\ \ldots ,\ 2,\ 2,\ 2]^{\rm T}, \quad & \quad V(2n+2) = [1,\ 2,\ 2,\ \ldots ,\ 2,\ 3,\ 2] \\
  U(2n+3) = [2,\ 2,\ 2,\ \ldots ,\ 2,\ 3,\ 2]^{\rm T}, \quad & \quad V(2n+3) = [2,\ 2,\ 2,\ \ldots ,\ 2,\ 3,\ 2] \\
  U(2n+4) = [2,\ 2,\ 2,\ \ldots ,\ 2,\ 4,\ 2]^{\rm T}, \quad & \quad V(2n+4) = [2,\ 2,\ 2,\ \ldots ,\ 2,\ 3,\ 3] \\
  U(2n+5) = [2,\ 2,\ 2,\ \ldots ,\ 2,\ 4,\ 3]^{\rm T}, \quad & \quad V(2n+5) = [3,\ 2,\ 2,\ \ldots ,\ 2,\ 3,\ 3] \\
  \ldots, \quad & \quad \ldots \\
  U(3n+3) = [3,\ 3,\ 3,\ \ldots ,\ 3,\ 2,\ 4,\ 3]^{\rm T}, \quad & \quad V(3n+3) = [3,\ 3,\ 3,\ \ldots ,\ 3,\ 3,\ 3,\ 3] \\
  U(3n+4) = [3,\ 3,\ 3,\ \ldots ,\ 3,\ 3,\ 4,\ 3]^{\rm T}, \quad & \quad V(3n+4) = [3,\ 3,\ 3,\ \ldots ,\ 3,\ 4,\ 3,\ 3] \\
  U(3n+5) = [3,\ 3,\ 3,\ \ldots ,\ 3,\ 4,\ 4,\ 3]^{\rm T}, \quad & \quad V(3n+5) = [3,\ 3,\ 3,\ \ldots ,\ 3,\ 4,\ 3,\ 4] \\
  U(3n+6) = [3,\ 3,\ 3,\ \ldots ,\ 3,\ 4,\ 4,\ 4]^{\rm T}, \quad & \quad V(3n+6) = [4,\ 3,\ 3,\ \ldots ,\ 3,\ 4,\ 3,\ 4] \\
  \ldots, \quad & \quad \ldots \\
  U(4n+3) = [4,\ 4,\ 4,\ \ldots ,\ 4,\ 3,\ 4,\ 4,\ 4]^{\rm T}, \quad & \quad V(4n+3) = [4,\ 4,\ 4,\ \ldots ,\ 4,\ 4,\ 4,\ 3,\ 4] \\
  U(4n+4) = [4,\ 4,\ 4,\ \ldots ,\ 4,\ 4,\ 4,\ 4,\ 4]^{\rm T}, \quad & \quad V(4n+4) = [4,\ 4,\ 4,\ \ldots ,\ 4,\ 5,\ 4,\ 3,\ 4] \\
  \ldots, \quad & \quad \ldots.
  \end{aligned}} \end{equation*} \caption{The padding dynamics for $I_{n+1}$.} \label{fig:padding dynamics for general n}
  \end{figure*}

\end{document}